\documentclass[11pt,letterpaper]{JHEP}
\title{Finite Noncommutative Chern-Simons with a Wilson Line and
the Quantum Hall Effect}
\author{Bogdan Morariu\thanks{morariu@summit.rockefeller.edu} 
\\Department of Physics, Rockefeller University   \\
        New York, NY 10021}
\author{Alexios P. Polychronakos\thanks{poly@theorfys.uu.se} \thanks{On 
leave from
Theoretical Physics Dept., Uppsala University,
  Sweden and Physics Dept., University of Ioannina, Greece}
\\Department of 
Physics, Rockefeller University   \\
        New York, NY 10021}
\preprint{ RU-01-9-B\\ CCNY-HEP 01/04}

\abstract{We present a finite dimensional matrix model associated to
the 
noncommutative
Chern-Simons theory, obtained by inserting a \mbox{Wilson} line. For a
specific choice
of the representation of the Wilson line the model is equivalent to
 minimal modification of the matrix model which is compatible with
finite dimensional matrices, and was introduced previously to study
droplets of quantum Hall fluid. For other representations we obtain
generalizations corresponding to regularized ${\rm U}(n)$ Chern-Simons
theories, representing multilayered quantum Hall fluids.}


\usepackage{amsmath,amssymb}

\def\11{\mbox{$1$}}
\newcommand{\rref}[1]{(\ref{#1})}
\newcommand{\beqn}{\begin{equation}}
\newcommand{\eeqn}{\end{equation}}
\newcommand{\beqarr}{\begin{eqnarray}}
\newcommand{\eeqarr}{\end{eqnarray}}

\newcommand{\matc}{\begin{array}{c}}
\newcommand{\matcc}{\begin{array}{cc}}
\newcommand{\matccc}{\begin{array}{ccc}}
\newcommand{\matcccc}{\begin{array}{cccc}}
\newcommand{\emat}{\end{array}}

\newcommand{\IH}{\relax{\rm I\kern-.18em H}}
\newcommand{\IR}{\relax{\rm I\kern-.18em R}}
\newcommand{\IK}{\relax{\rm I\kern-.18em K}}
\newcommand{\II}{\hbox{\rm 1\kern-.28em I}}
\newcommand{\Is}{\relax{\rm 1\kern-.35em 1}}

\begin{document}

%\newpage
%\renewcommand{\thepage}{\arabic{page}}

%\setcounter{page}{1}
%\setcounter{footnote}{0}

% main text is here
\section{Introduction}
\label{Intro}

Recently, Susskind~\cite{Susskind:2001fb} proposed that the
coordinates of electrons  moving in a strong magnetic field $B$
be described by matrices, similar to the description of $D0$ brane
coordinates. The corresponding lowest Landau level action,
consisting of only the magnetic term \cite{DuJT}, becomes a noncommutative
version of the Chern-Simons action, which has found 
numerous applications in physics  \cite{DJT}. In the present context, it
describes Laughlin fractional quantum Hall states \cite{Laug}, 
the Chern-Simons level becoming the inverse filling fraction.
For a recent introduction to the quantum Hall effect see~\cite{SMGirvin}.

The above Chern-Simons theory can describe only an infinite
number of electrons. In a previous paper, one of us (A.P.)
proposed a regularized version of the noncommutative theory 
on the plane in the form of a Chern-Simons matrix model with 
boundary terms \cite{Polychronakos:2001mi}. 
This model describes a system of
finitely many electrons (a quantum Hall `droplet') and reproduces 
all the relevant physics of the
finite Laughlin states, such as boundary excitations \cite{Wen,IKS},
quantization of the filling fraction (see also \cite{NP,BLP})
and quantization of the charge of quasiparticles (fractional holes). 
An extension of this model for electrons on a cylinder, involving
unitary matrices, is also introduced in \cite{PolQHC}.
An explicit, but non-unitary, mapping between the states of
the matrix model and Laughlin states was presented in 
\cite{Hellerman:2001rj}, while possible wavefunction 
mappings were explored in \cite{KS}.

An essential ingredient of the above models is the so-called
`boundary' term. Its role is to absorb the anomaly of the theory
and allow a finite matrix representation of the Gauss' law. In practice,
it `feeds' the representation of the gauge group ${\rm U}(N)$ carried by
the matrix coordinates of the model. This representation is fixed 
by the filling fraction and 
determines the physics of the electrons; it is the $(n-1)N$-fold
symmetric representation of ${\rm U}(N)$ \cite{PolMM}, with the integer
$n = \nu^{-1}$ representing the inverse filling fraction.

A natural question is whether other boundary terms could have been
chosen and, correspondingly, different representations for ${\rm U}(N)$.
To approach this, we observe that the boundary term essentially
corresponds to a Wilson line operator (see also \cite{PolQHC})
and that a natural expression for such Wilson lines is through
a first order gauge invariant action on the group 
manifold~\cite{Balachandran:1983pc}.
We therefore propose
to use such a Wilson line action as the boundary 
term of the matrix model and study its classical and quantum
structure.

The layout of this paper is as follows.
In Section~\ref{Review} we review the noncommutative Chern-Simons
and its finite dimensional cousin. Section~\ref{WL} describes the
proposed generalization of the action 
of~\cite{Polychronakos:2001mi}. Quantization of this action is
equivalent to the insertion of a Wilson line.
In Section~\ref{QE} we show how our proposal
relates to that of~\cite{Polychronakos:2001mi}. We discuss both the 
classical and quantum equivalence. In the
last section  we discuss the quantum Hall interpretation and possible
applications of the model.
Finally, in order to fix our conventions and to make the paper relatively
self-contained, we have included a brief description of the
quantization of the symplectic form on coadjoint orbits in an appendix.

\section{Review of Noncommutative Chern-Simons in the Quantum Hall Effect}
\label{Review}

In Susskind's proposal, the dynamics of a system of $N$ 
electrons  in the large $N$ limit is described by the action
\begin{equation}
{\cal S}_{NCCS}
=
\int dt\frac{B}{2} {\rm \,Tr\,}
\{
\epsilon_{ij}(\dot{X}^i+i[A_0,X^i]) X^j + 2\theta A_0
\}~,
\label{CSaction}
\end{equation}
where $\theta$ is related to the average planar electron density
$\rho_0= 1/2\pi \theta$. The subscript $NCCS$ stands for noncommutative
Chern-Simons and will be explained shortly.
Similar actions first appeared in matrix Chern-Simons theory as a
possible approach to the fundamental formulation of 
M-theory~\cite{Smolin:1998ai}.

Gauss' law obtained by varying $A_0$ in the action~\rref{CSaction}
takes the Heisenberg algebra form
\begin{equation}
[X^1,X^2]=i\theta ~.
\label{GL}
\end{equation}
Let $X^i= y^i$ be the solution of the Gauss' law equation~\rref{GL}
corresponding to $y^1=x$ and $y^2=\theta p$ where $x$ and $p$ are the
matrices representing the position and momentum operators
in the harmonic oscillator basis.
If we insert $X^i= y^i+\epsilon^{ij}\theta A_j$ into the the
action~\rref{CSaction}, where $A_i$ parameterizes perturbations
around the $y^i$ solution, one obtains the ${\rm U}(1)$ 
noncommutative Chern-Simons
action in the operator language \cite{PolCS}
\[
{\cal S}_{NCCS}=
\frac{k}{4\pi} \int dt \, \epsilon^{\mu \nu \xi}\, 2\theta\,{\rm Tr}
\left(
A_{\mu} [\partial_{\nu}, A_{\xi}]+\frac{2}{3} A_{\mu} A_{\nu} A_{\xi}
\right)~.
\]
Here $\partial_i \equiv i\theta^{-1}\epsilon_{ij} y^i$ and is regarded
as a (matrix) operator. The level $k$ of the Chern-Simons action equals
(classically) the
inverse filling fraction $\nu^{-1}$ where the filling fraction $\nu$
is defined as
\[
\nu=
\frac{2\pi\rho_0}{B}~.
\]

Taking the trace of~\rref{GL} we see that Gauss' law is only compatible
with infinite dimensional matrices.
In order to describe a finite number of electrons, one of the
authors~\cite{Polychronakos:2001mi}, proposed a minimal addition to the
action~\rref{CSaction} to make it compatible with finite dimensional
matrices. The proposed action is
\begin{equation}
{\cal S}=
{\cal S}_{NCCS}+{\cal S}_{\Psi}+{\cal S}_{\omega}~,
\label{PCS}
\end{equation}
where
\beqarr
{\cal S}_{\Psi}&=&    \int dt  \Psi^{\dagger}(i\dot{\Psi}-A_0\Psi)~,
\label{SPsi}\\
{\cal S}_{\omega}&= - &\int dt B {\rm \,Tr\,}\{\omega (X^i)^2\}~.
\label{Somega}
\eeqarr
The action~\rref{SPsi} involves a complex bosonic N-vector $\Psi$ such
that we have a ${\rm U}(N)$ gauge invariance
\[
\Psi \rightarrow U\Psi~,~~
X^i  \rightarrow U X^i U^{-1}~,~~
\partial_t+i A^0  \rightarrow U (\partial_t+i A_0) U^{-1}~.
\]
The term~\rref{Somega} is just a spatial regulator, a harmonic
potential whose role is to keep the electrons close to the origin. One
can take $\omega$ arbitrarily small.

The $A_0$ equation of motion is now modified and reads
\begin{equation}
iB \, [X^1,X^2]-\Psi\Psi^{\dagger}+B\theta=0~.
\label{ModGL}
\end{equation}
The trace of~\rref{ModGL} then simply implies
\begin{equation}
\Psi^{\dagger}\Psi =Nk~,
\label{U1}
\end{equation}
which is compatible with finite dimensional matrices. It was shown
in~\cite{Polychronakos:2001mi}, using the equivalence  to the
Calogero model, that upon quantization this
system describes Laughlin states
for the Quantum Hall effect of $N$ electrons
at fractional filling fraction $\nu=1/(k+1)$.
For a recent discussion of the equivalence of~\rref{PCS} and the Calogero
model see~\cite{Polychronakos:1999sx} and references there.
In the $A_0=0$ gauge, one can first quantize the system ignoring the
constraint~\rref{ModGL} and then impose it on physical states.
In the absence of the constraint we have a system of free harmonic
oscillators, some coming from the matrix elements of
$X^i$ and some from the components of $\Psi$.
Gauss' law~\rref{ModGL} implies
that physical states are invariant under ${\rm U}(N)$
gauge transformations\footnote{We have included the 
background charge $B\theta$\,.}.
A group theoretical analysis of this physical state condition
was used in~\cite{Hellerman:2001rj} to find a more direct
mapping to Laughlin states.

\section{The Wilson Line Action}
\label{WL}
In this section we propose a generalization of the action~\rref{PCS}. We
propose to replace ${\cal S}_{\Psi}$ with
with
\begin{equation}
{\cal S}_{g}
=
\int dt\, {\rm Tr} \left[i \lambda g^{-1} (\partial_t+iA_0)
g\right]~,
\label{Sg}
\end{equation}
where $g$ is valued in the ${\rm U}(N)$ group and $\lambda$ is an
arbitrary hermitian matrix~\cite{Balachandran:1983pc}.
Without loss of generality $\lambda$ can
be taken to be diagonal.
Let $H$ denote the stability group of $\lambda$ under the
adjoint action of ${\rm U}(N)$. For generic $\lambda$ the subgroup $H$
is just the diagonal Cartan subgroup. 

This system is gauge invariant under the 
transformations $g'=gh$, where $h$ is time dependent and $H$ valued.
In fact, the action~\rref{Sg} is invariant under infinitesimal transformations.
For large gauge transformations the action changes by an integer multiple of
$2\pi$ only if $\lambda$ is quantized. 
The gauge invariant degrees of freedom are points on the coadjoint
orbit of $G$ passing through $\lambda$\,. These are symplectic
leaves, and it is a well known fact~\cite{BKAAK} that 
their quantization gives the unitary representations of the Lie
group $G$\,. 
The quantization of this system
for $g$ valued in a simple Lie group $G$ is reviewed in the
appendix. The method there, is similar in spirit to the one
used  in~\cite{Alexanian:2001qj}. See 
also~\cite{Morariu:1999vu} for a path integral quantization, using Darboux 
coordinates.
Here we only explain the necessary modifications for
\[
{\rm U}(N)
\cong\left(
{\rm SU}(N)\times {\rm U}(1)\right)\, /\, {\mathbb Z}_N~.
\]

First let us obtain the quantization  of $\lambda$. Let the Cartan
generators $H_i$ of ${\rm SU}(N)$ be
\begin{equation}
H_i={\rm diag}(0,\ldots,0,1,-1,\ldots,0)~,~~i=1,\ldots,N-1~,
\label{HI}
\end{equation}
where the only nonvanishing entries are on the $i$ and $i+1$ positions.
To this we append $H_N={\mathbb I}$\, so that we have a complete set in the
Cartan subalgebra of ${\rm u}(N)$.
Similarly, let the fundamental weights of ${\rm su}(N)$ be given
by\footnote{Since the trace is nondegenerate we will identify the Lie
algebra with its dual. The same can be done for the Cartan subalgebra
and its dual.}
\[
\mu^i=\frac{1}{N}\,
{\rm diag}(N-i,\ldots,N-i,-i,\ldots,-i)~,~~i=1,\ldots,N-1~,
\]
where the first $i$ and the last $N-i$ diagonal entries of $\mu^i$ are equal.
To this set we add $\mu^N= 1/N\,{\mathbb I}$  such that we now have
${\rm Tr}\,
(H_i\, \mu^j)=\delta_i^j,~i=1,\ldots,N$.
Using this we can write $\lambda =-n_i \mu^i$ where
$n_i=-{\rm Tr}(\lambda H_i)$.
Consider now a large gauge transformation of the form
\[
h(t)= e^{2\pi i H_i \xi^i(t)}~,
\]
where $\xi^i$ are real functions of time
such that $h(t)={\mathbb I}$ for large positive or negative
times. This periodicity implies that
\begin{equation}
H_i \Delta \xi^i ={\rm diag}(k_1,\ldots,k_N)~,
\label{Diagonal}
\end{equation}
where the diagonal elements $k_i$ must be integers. Under the gauge
transformation  $h(t)$ the Lagrangian changes by a total derivatives
which can be integrated to give the following change of the action
\[
\Delta {\cal S}_{g}
=
2\pi
{\rm Tr}
\left[-
\lambda(H_i \Delta \xi^i)\right]
=2\pi
n_i \Delta\xi^i~.
\]
The requirement that the action only changes by an integer multiple of
$2\pi$ together with~\rref{Diagonal} implies that all the diagonal
elements of $\lambda$ must be integers. Then $n_i=-{\rm Tr}(\lambda H_i)$
are also integers.
For arbitrary $h\in H$, one can show that  $\lambda_i$\,, the diagonal
matrix elements of $\lambda$ satisfy 
\[
\lambda_i=-
\frac{1}{N}
\left[
\sum_{i=1}^{N-1} (N-i)n_i +n_N\right]+0\,{\rm mod}(1)~.
\]
Thus, to be free of
global gauge anomalies $n_i$ must all be integers and satisfy
\begin{equation}
\sum_{i=1}^{N-1} (N-i) n_i +n_N=0\,{\rm mod}(N)~.
\label{HWC}
\end{equation}

The quantization can be now be performed similarly to the
treatment in the appendix of a simple Lie group. Here we will be brief
and only outline the steps (see the appendix for details).
The same dynamics can be obtained from  a nondegenerate first order
action
\begin{equation}
{\cal S}=
\int dt\, {\rm Tr} \left[i p g^{-1} \dot{g} - gpg^{-1}A_0\right]~,
\label{NCA}
\end{equation}
together with the constraint
\begin{equation}
p-\lambda =0~.
\label{CONST}
\end{equation}
The symplectic structure obtained from the first term in~\rref{NCA}
is the standard one on the cotangent bundle of ${\rm U}(N)$.
The Hilbert space of the unconstrained system is given
by square integrable functions on ${\rm U}(N)$ whose harmonic expansion
is given by Peter-Weyl theorem. Just as in the appendix, one then
imposes only the constraints~\rref{CONST}
associated with the Cartan and positive
roots generators. The physical Hilbert space is then finite
dimensional and provides an irreducible representation of ${\rm U}(N)$.
It is the representation whose {\em lowest} weight is given by $\lambda$.
Note that the first term on the left hand side of~\rref{HWC} is just
the number of boxes in the Young tableau of the ${\rm SU}(N)$
representation, while the second is the ${\rm U}(1)$ charge.

Finally we can now explain the title of this section. The path
integral over $g$ in the action~\rref{Sg} gives a Wilson line in the
irreducible representation discussed above
\begin{equation}
\int dg \,e^{i{\cal S}_{g}}
=
{\rm P} e^{i\int dt \, A_0^a(t)t_a}~,
\label{timeorder}
\end{equation}
where $\rm P$ denotes path ordering.
Both the right and the left hand side of~\rref{timeorder} are
understood as operators acting in the physical Hilbert space.
This can be seen as follows. Using the notation in the appendix we can write
\[
{\rm Tr} (\lambda g A_0 g^{-1})
=
\tilde{p}_a A_0^a~.
\]
Upon quantization as can be seen from~\rref{tildep},
$-\tilde{p}_a$ becomes the operator acting on the
Hilbert space as $t_a$. Since the relation between operators and path
integral also involves time ordering, which is path ordering in this
case, we obtain the desired result.

Gauss' law for the full action
${\cal S}= {\cal S}_{NCCS}+{\cal S}_{g}+ {\cal S}_{\omega}$
now reads
\begin{equation}
iB \, [X^1,X^2]-g\lambda g^{-1}+B\theta=0~,
\label{ModGLg}
\end{equation}
and the trace of~\rref{ModGLg} implies a constraint on the trace of $\lambda$
\begin{equation}
{\rm Tr}(\lambda) =Nk~. \nonumber
\end{equation}

\section{Relation to Boundary Fields}
\label{QE}

In this section we will demonstrate the correspondence of this model
with the previous model with boundary $\Psi$ terms. Specifically, we
will show that the Wilson line action involving $g$ is a particular sector
of a version of the previous model involving $N$ different boundary 
fields $\Psi$.

First let us discuss the classical equivalence. The ${\rm U}(N)$ matrix $g$
can be written in terms of its columns $\psi_j$, $j=1,2, \dots N$. Since
$g^{-1} = g^\dagger$, the action ${\cal S}_{g}$ becomes
\begin{equation}
{\cal S}_{g}
=
\int dt\, \sum_j \left[i\lambda_j \, \psi_j^\dagger (\partial_t +iA_0)
\psi_j \right]~.
\label{Spsij}
\end{equation}
We have the sum of $N$ decoupled actions. The vectors $\psi_j$
are still coupled through the relations
\begin{equation}
\psi_j^\dagger \psi^{}_k = \delta_{jk}~,
\label{psic}
\end{equation}
implied by the unitarity of $g$. Upon imposing Gauss' law, however,
and putting $A_0 = 0$ the equations of motion for $g$ are
\begin{equation}
{[} \lambda , g^{-1} {\dot g} {]} = 0~,
\end{equation}
and in terms of $\psi_j$ they become
\begin{equation}
(\lambda_j - \lambda_k ) \, \psi_j^\dagger {\dot \psi}^{}_k = 0 
~~~{\rm (no~sum~in}~j,k{\rm )}~.
\label{eompsi}
\end{equation}
For generic values of $\lambda_j$ the above equations imply
that ${\dot \psi}_j = 0$ is orthogonal to all other $\psi_k$, and therefore
${\dot \psi}_j \sim \psi_j$. Since $\psi_j$ is normalized, this means that only
its phase can vary. The Wilson line 
action (\ref{Sg}), on the other hand, has an
additional ${\rm U}(1)^N$ gauge invariance, corresponding to right-multiplication of
$g$ by an arbitrary diagonal unitary matrix. This transformation is exactly the
redefinition of the phases of $\psi_j$. So the above motion of $\psi_j$
corresponds to a ${\rm U}(1)^N$ gauge transformation and, as a gauge choice, 
we can take ${\dot \psi}_j = 0$. These are the same equations of motion that
we would have derived from the action (\ref{Spsij}), ignoring the
constraints (\ref{psic}). The constraints, then, can be imposed as
initial data for the $\psi_j$.

We can go one step further and redefine the $\psi_j$ to absorb $\lambda_j$.
We distinguish between the cases $\lambda_j <0$ and $\lambda_j >0$.
We define
\begin{equation}
\Psi_j = \sqrt{| \lambda_j |} \, \psi_j~.
\end{equation}
Assuming that there are $n$ positive and $N-n$ negative $\lambda$'s
the action becomes
\begin{equation}
{\cal S}_{g}
=
\int dt\, \sum_{j=1}^n \left[ \Psi_j^\dagger (i\partial_t -A_0)
\Psi_j \right] + \sum_{j=n+1}^N \left[ \Psi_j^\dagger (-i\partial_t +A_0)
\Psi_j \right]~.
\label{SPsij}
\end{equation}
This is identical to the original action ${\cal S}_\Psi$, where, now, we
have introduced a multiplet of boundary fields $\Psi_j$, with $n$ of them
transforming under the fundamental of the gauge group and $N-n$
transforming under the anti-fundamental representation. We must further
choose as initial conditions
\begin{equation}
\Psi_j^\dagger \Psi_k^{} = |\lambda_j |  \, \delta_{jk}~.
\label{orth}
\end{equation}
So we have traded $\lambda$ for the initial conditions of the lengths of
$\Psi_j$, and we have the additional condition of orthogonality between
the different $\Psi_j$. This last requirement is not a restriction. The generator
of gauge transformations for the $\Psi_j$, which enters Gauss' law, is
\begin{equation}
G_\Psi = \sum_{j=1}^n \Psi_j  \Psi_j^\dagger - 
\sum_{j=n+1}^N \Psi_j \Psi_j^\dagger~.
\label{GP}
\end{equation}
This is a hermitian matrix which obviously projects in the space spanned 
by $\Psi_j$, so it can be diagonalized by a unitary transformation in this
space. This amounts to a linear redefinition of the $\Psi_j$ such that they be
orthogonal to each other. These redefined $\Psi_j$ satisfy (\ref{orth}).

If any of the $\lambda_j$ are equal to each other then the equations
of motion (\ref{eompsi}) do not imply that ${\dot \psi}_j$ is proportional
to $\psi_j$, for the corresponding values of $j$, but rather to an arbitrary
linear combination of these $\psi_j$'s. In this case, however, the Wilson 
line action has an enhanced gauge invariance under 
right-multiplications of $g$ by unitary matrices in the corresponding 
subspace. The equations of motion then allow an arbitrary motion in
this subspace, which becomes a gauge transformation.
It is consistent, therefore, to also impose ${\dot \psi}_j = 0$ in this
case, which amounts to a gauge choice. The gauge generator
$G_\Psi$ in (\ref{GP}) in this case will have a degenerate subspace
for the corresponding $j$, and the gauge arbitrariness corresponds to the
freedom of rotating the $\Psi_j$ in this subspace. 

We conclude that in all cases the Wilson line action ${\cal S}_{g}$
is essentially equivalent to the action ${\cal S}_\Psi$ for $N$ independent
boundary fields $\Psi_j$ transforming in the (anti)fundamental of
${\rm U}(N)$, after fixing the initial conditions and choosing 
the basis where the matrix generator $G_\Psi$ in the
Gauss' law is already diagonalized. The corresponding theories
with many boundary fields are similar to the models in \cite{MP,AJ}
which, upon proper reduction, lead to spin-Calogero models.

Quantum mechanically a similar picture emerges. In the theory with
$n$ boundary fields in the fundamental and $N-n$ boundary fields
in the antifundamental the components of the fields $(\Psi_j )_k$ for $j
\le n$, and $(\Psi_j )_k^\dagger$ for $j > n$ (where $\dagger$
now stands only for operator, not matrix, hermitian conjugation)
become oscillator
annihilation operators (note the opposite sign of the canonical term in
(\ref{SPsij}) for $j>n$).  The boundary gauge generators are the sum of
$N$ independent oscillator realizations of the generators of ${\rm U}(N)$.
Renaming $(\Phi_j)_k = (\Psi_j)_k^\dagger$
 we have the ordering
\begin{equation}
G_\Psi^a = \sum_{k=1}^n (\Psi_k )_i^\dagger T_{ij}^a  (\Psi_k )_j - 
\sum_{k=n+1}^N (\Phi_k )_i^\dagger T_{ji}^a (\Phi_k )_j~.
\label{GPop}
\end{equation}
The Fock spaces of the $\Psi,\Phi$ embeds the tensor product of
the representations of each oscillator realization; for $j\leq n$ the
oscillator $\Psi_j$ reproduces all totally symmetric irreps of ${\rm SU}(N)$
(all irreps with a single row in their Young tableau), 
with ${\rm U}(1)$ charge equal to the total number operator.
For $j>n$ the oscillator $\Phi_j$ reproduces the conjugates
of the above representations (irreps with $N-1$ rows of equal length). 
The tensor product of $N$ such irreps of all possible lengths contains
all representations of ${\rm SU}(N)$. Therefore, the model with $N$ boundary fields
allows for the most general representation of the Gauss' law generator.
Picking a particular irrep is the quantum mechanical analog of fixing
initial conditions and diagonalizing the classical gauge generator matrix.
The action ${\cal S}_{g}$, on the other hand, also reproduces an arbitrary
representation of the Gauss' law generator, upon picking the appropriate
$\lambda$. 

We conclude that the two theories are essentially equivalent, with the 
difference that ${\cal S}_{g}$ picks an irreducible component for the 
Gauss' law generator, that is, a particular 
sector of the ${\cal S}_\Psi$ theory.
The original single-$\Psi$ theory only has one sector and it is
completely reproduced by the Wilson line model with 
all but a single of the $\lambda_j$'s vanishing:
\begin{equation}
\lambda=-Nk\mu^{N-1}+Nk\mu^N
={\rm diag}(0,\ldots,0,Nk) ~.\label{Lambda}
\end{equation}

Notice that the ${\rm U}(1)$ charge of the representation, determined
by ${\rm tr} \lambda$, corresponds to the total number operator of the
oscillators. By adding such a ${\rm U}(1)$ part we could render all
$\lambda_j$ positive and dispense with the anti-fundamental
fields in the $\Psi$ representation. We stress, however, that the
total ${\rm U}(1)$ charge essentially determines the noncommutativity
parameter $\theta$ (the filling fractions) through the Gauss' law.
Including, therefore, some antifundamental fields is crucial is we
wish to reproduce all irreps of ${\rm SU}(N)$ for a fixed value of $\theta$.

\section{Discussion}

The proposed generalization allows for the maximal flexibility in the
choice of the representation for the Gauss' law for the matrix 
coordinates $X_i$. The Wilson line ($g$) representation is particularly
convenient in that it picks a single irreducible representation.
The boundary field ($\Psi_j$) representation, on the other hand, is
most suited to discuss the physics of the model.

One obvious application of the many-$\Psi$ model would be to
describe many layers of quantum Hall fluids and/or fluids with spin.
Intuitively, the presence of many boundary fields creates many boundaries
for the quantum Hall droplet which, then, decomposes into many layers.
The total number of electrons in all layers is always $N$.

As an example, consider the case of two fields $\Psi_{1,2}$, both
in the fundamental,  satisfying
\begin{equation}
\Psi_i^\dagger \Psi_j = k N_i \delta_{ij} ~,~~~ N_1 + N_2 = N~,
\end{equation}
where $N_{1,2}$ are integers. Then the Gauss' law implies
\begin{equation}
iB[X^1,X^2] + k \, {\rm diag} \left( 1,\dots 1-N_1, 1,\dots 1-N_2 \right)=0~.
\label{Gautwo}
\end{equation}
We have chosen a basis in which the Gauss' law is diagonal, 
with the entry \mbox{$1-N_1$} appearing in the position $N_1$ in the diagonal.
In this way the trace of the first $N_1$ elements of the above matrix
vanishes. So (\ref{Gautwo}) admits block-diagonal solutions for
$X_1$ and $X_2$, each representing a quantum Hall droplet with
$N_1$ or $N_2$ electrons. The two droplets can obviously overlap.
So this model can describe different quantum Hall layers as well
as their interactions (non-block diagonal solutions). The generalization
to $n$ layers is straightforward. The layers can, equivalently, be
viewed as spin components for the electrons. 
The relation of the $n$-component model
to the ${\rm SU}(n)$-spin Calogero model \cite{MP} further enhances the
likelihood of such a correspondence, although the details are 
yet to be worked out.

Another obvious application of the proposed model is as a
regularization of ${\rm U}(n)$ noncommutative Chern-Simons theory.
For the infinite plane case the Gauss' law (\ref{GL}) can admit
reducible representations corresponding to the direct sum of
$n$ Heisenberg representations. Perturbations around
such a solution would give rise to ${\rm U}(n)$ NCCS action.
The single-boundary matrix model, on the other hand, has
a ground state corresponding to a single layer which reproduces
${\rm U}(1)$ NCCS theory. 
By choosing, however, $n$ boundary fields in the
fundamental and taking them to be orthogonal and of norm 
squared $kN/n$ ($N$ should be a multiple of $n$) we have
a situation analogous to the one above, which admits as
ground state a block-diagonal configuration which is the
direct sum of $n$ ground states of size $N/n$ each.
Clearly there is an extra ${\rm U}(n)$ symmetry mixing the $n$
components.
Perturbations around this configuration would give a regularized
version of ${\rm U}(n)$ theory.

In conclusion, the extended model proposed here has many
potential applications both in noncommutative gauge theory
and in the quantum Hall context, which are the topic of
further investigation.

\acknowledgments
%\section*{Acknowledgments}
We would like to thank Klaus~Bering, Dimitra Karabali,
Parameswaran~Nair, Bunji~Sakita and  Lenny Susskind for illuminating 
discussions. A.P. would also like to thank the Physics Department of
Columbia University for hospitality during part of this work. This
work was supported in 
part by the U.S.~Department of Energy
under Contract Number DE-FG02-91ER40651-TASK B.

\appendix
\section{Coadjoint orbits quantization}

In this appendix we describe the quantization of the system defined by
the action
\begin{equation}
{\cal S}=
\int dt\, {\rm Tr} \left[i \lambda g^{-1} (\partial_t+iA_0) g\right]~,
\label{action}
\end{equation}
where $g$ is valued in the group $G$ which we assume to be a
simple Lie group. See~\cite{Balachandran:1983pc} and references there, and
for a recent treatment see~\cite{Alexanian:2001qj}. 
The normalization of the trace is chosen such that
the length square of a long root equals two. For $G={\rm SU}(N)$ this
reduces to the matrix trace in the defining representation.
In~\rref{action} $\lambda$ is a constant weight.

The action~\rref{action} is equivalent to the action
\begin{equation}
{\cal S}=
\int dt\, {\rm Tr} \left[i p g^{-1} \dot{g} - gpg^{-1}A_0\right]~,
\label{Cotangent}
\end{equation}
together with the constraint
\begin{equation}
p-\lambda =0~.
\label{Const}
\end{equation}
The first term in the action~\rref{Cotangent} gives the standard
symplectic form on the cotangent bundle of the group $G$
\[
\omega =
d\, {\rm Tr}(i pg^{-1}dg)~,
\]
from which we can derive the Poisson bracket
\[
i\,\{f,h\}=
{\rm Tr}
\left(
\frac{\partial f}{\partial g^{T}}
g
\frac{\partial h}{\partial p^{T}}
-
\frac{\partial h}{\partial g^{T}}
g
\frac{\partial f}{\partial p^{T}}
+
\frac{\partial f}{\partial p^{T}}
p
\frac{\partial h}{\partial p^{T}}
-
\frac{\partial h}{\partial p^{T}}
p
\frac{\partial f}{\partial p^{T}}
\right)~.
\]
Let $p_a = -{\rm tr}(t_a p)$ and $g_{MN}$ be coordinates on the
cotangent bundle. They have the following Poisson brackets
\[
i\{p_a,p_b\}=
i f_{ab}^{~~c} \,\,p_c~,~~
i\{p_a, g_{MN}\}=
(g t_a)_{MN}~,
\]
which upon quantization become
\[
[p_a,p_b]=
i f_{ab}^{~~c} \,\,p_c~,~~
[p_a, g_{MN}]=
(g t_a)_{MN}~.
\]
Let $\tilde{p}= gpg^{-1}$ designate the matrix in front of $A_0$ in the
action~\rref{Cotangent}. Then the operators
$\tilde{p}_a={\rm tr}( t_a \tilde{p})$ commute with $p_a$ and satisfy
\begin{equation}
[\tilde{p}_a,\tilde{p}_b]=
i f_{ab}^{~~c} \,\,\tilde{p}_c~,~~
[\tilde{p}_a, g_{MN}]=
(-t_a g)_{MN}~.
\label{tildep}
\end{equation}

We can represent the Hilbert space $\cal H$ of the unconstrained system using
square integrable functions on the group manifold.
By Peter-Weyl theorem any function on the
group has the decomposition
\begin{equation}
\psi(g)=
\sum_R \sum_{\alpha \beta} c_R^{\alpha \beta} \,R_{\alpha \beta}(g)~,
\label{PW}
\end{equation}
where $R_{\alpha \beta}(g)$ are the matrix elements of the $R$
representation of $G$.  They satisfy the following orthogonality
conditions
\[
\int\, dg \, R_{\alpha \beta}(g)\, \bar{R}'_{\rho \sigma}(g)
=
\frac{1}{d_R}
\delta_{RR'} \,\delta_{\alpha\rho}\, \delta_{\beta \sigma}~,
\]
where $dg$ is the the Haar measure and
$d_R$ is the dimension of the $R$ representation.

The operators
$p_a$ and $\tilde{p}_a$ act on this Hilbert space generating an
action of $G_l\times G_r$, where we used a subscript to distinguish
the two $G$ factors.
Under the
action of $(g_l^{~},g_r) \in G_l\times G_r$ we have the following
transformation
\begin{equation}
\psi(g)
\rightarrow
\psi'(g)
=
\psi(g_l^{-1}\, g\, g_r)~.
\label{LRaction}
\end{equation}
In particular $R_{\alpha \beta}(g)$ transforms as
\[
R_{\alpha \beta}(g)
\rightarrow
R_{\rho \sigma}(g) \bar{R}_{\rho\alpha}(g_l^{~}) R_{\sigma \beta}(g_r)~,
\]
that is, the index $\beta$ transforms in the representation $R$ while the
index $\alpha$ transforms in the representation $\bar{R}$.
We can reinterpret~\rref{PW} as the decomposition of the Hilbert space
$\cal H$  under the action~\rref{LRaction}
into the following sum of irreducible representations
\begin{equation}
{\cal H}
\cong
\bigoplus_{R}
V_{\bar{R}}\otimes V_{R}~.
\label{Hilbert}
\end{equation}
Here $V_R$ is the vector space in which the representation $R$ acts,
$\bar{R}$ is the complex conjugate representation and the sum is
over all the inequivalent unitary irreducible representations of
$G$. If the states $|\alpha,R\rangle$, $\alpha=1,\ldots,d_R$ form a basis of
$V_R$ and $|\overline{\alpha,R}\rangle$ a basis of $V_{\bar{R}}$,
the isomorphism is given by
\[
R_{\alpha\beta}(g)
\rightarrow
|\overline{\alpha,R}\rangle\otimes |\beta,R\rangle~.
\]

Let us now consider the constraint~\rref{Const}. Using the raising
and lowering generators $e_{\alpha}$ and  $e_{-\alpha}$, where $\alpha$
are positive roots (not to be confused with the index labeling the
components of representations), and $H_i$ forming a basis in the
Cartan subalgebra\footnote{Using the trace, $H_i$ can be identified
with the simple coroots $\alpha_i^{\vee}$.} we define
\beqarr
p_i~~~&=& -\,{\rm tr}(H_i\,p)~,~~i=1,\ldots,{\rm rank} (G)~,\nonumber \\
p_{\alpha}~~&=& -\,{\rm tr}(e_{\alpha} \,p)~,~~
p_{-\alpha}= -\,{\rm tr}(e_{-\alpha} \,p)~,~~\alpha > 0~.\nonumber
\eeqarr
Using ${\rm tr} (H_i\,\mu^j)=\alpha_i^{\vee}(\mu^j)  =\delta_i^j$
we can rewrite the constraint~\rref{Const} as
\begin{equation}
p_{i} - n_{i}=p_{\alpha}=p_{-\alpha} =0 ~,~~\alpha>0~,\label{Constranits}
\end{equation}
where $n_i= -{\rm tr} (H_i \lambda)$\,.
At the classical level, some of these constraints are first class and
some are second class. To see this, note that for some positive roots
$\alpha$ the Poisson bracket $\{p_{\alpha},p_{-\alpha}\}$ gives linear
combinations of the Cartan $p_i$ which can be nonvanishing due to the
constraints $p_i=n_i$.
At the quantum level we can not impose all the constrains
simultaneously. We can however use a Gupta-Bleuer type quantization and
require that physical states $\psi$ satisfy only
\beqarr
p_i\, \psi(g) &=& n_i\,\psi(g) \label{HW}~, \\
p_{\alpha}\,\psi(g)&=&0~,~~\alpha>0~. \nonumber
\eeqarr
These are the defining relations for a highest weight state
which only exists in the decomposition~\rref{Hilbert} if $n_i$ are
positive integers.
This is how the quantization of the weight $\lambda$ is obtained in
the Hamiltonian approach. Physical states must therefore have the form
\[
|\overline{\beta,R}\rangle\otimes |\beta_0,R\rangle~~~\beta=1,\ldots,d_R~,
\]
where $|\beta_0,R\rangle$ is the highest
weight state of weight $-\lambda$.
In the wave function on the group description we are restricted to
functions of the form
\[
\psi (g) =
\sum_{\beta} c^{\beta} R_{\beta \beta_{0}}(g)~.
\]
The physical Hilbert space is the irreducible representation of $G_l$  
whose complex conjugate representation highest weight is
$-\lambda$. Equivalently, the physical Hilbert space is the irreducible
representation whose {\em lowest} weight \mbox{is $\lambda$}.

\end{document}